\documentclass[preprintnumbers,amsmath,amssymb,floatfix,12pt,prd,superscriptaddress,nofootinbib]{revtex4}
\usepackage{graphicx}
\usepackage{epsfig}
\usepackage{bm}
\usepackage{amsfonts}
\usepackage{subfigure}

\def\bea{\begin{eqnarray}}
\def\eea{\end{eqnarray}}

\begin{document}
\begin{center}
\LARGE { \bf  Some aspects of  non-abelian gauge field inflation  model
  }
\end{center}
\begin{center}
{\bf M. R. Setare\footnote{rezakord@ipm.ir} \\  V. Kamali\footnote{vkamali1362@gmail.com}}\\
 { Department of Science, University of Kurdistan,\\
Sanandaj, IRAN.}
 \\
 \end{center}
\vskip 3cm

\begin{center}
{\bf{Abstract}}\\
In this paper we study  non-abelian gauge field inflation (gauge-flation) model in the context of intermediate and logamediate scenario. Important parameters of this model are presented for these two cases. Numerical study shows that the intermediate and logamediate gauge-flation are compatible with WMAP9 observational data. For intermediate inflation, where the cosmological scale factor expands as: $a(t)=a_0\exp(At^f)$ ($A>0, 0<f<1$), $N\geq 100$ cases lead to $0.02<r<0.11$ ($r$ is tensor-scalar ratio) and  $N\simeq50 $ case leads to $n_s\simeq 0.96$ ($n_s$ is spectral index). These constraints of perturbation parameters are agree with Planck and WMAP9 data. For logamediate model, where the scale factor expands as: $a(t)=a_0 \exp(a[\ln t]^{\lambda})$ ($\lambda >1, A>0$), $N\geq 200$ and $\lambda=2$ cases lead to interesting agreement with observational data.
 \end{center}
PACS Number(s): 98.80.Cq
\newpage

\section{Introduction}
The idea of inflationary universe presents the better description of the early phase of the universe \cite{1-m}. Some of the theoritical problems of the Hot Bing Bang (HBB) model, namely the numerical density of monopoles, homogeneity, flatness and the horizon problem will find explanations in the framework of the inflationary universe model.  Large scale structure (LSS) and cosmic microwave background (CMB) observational data denote the isotropic space-time at the background level \cite{2-m}. Inflation models which are proposed versus single or multi scalar field theories,  are isotropic and compatible with the observations of CMB and LSS. These  models have (non)standard kinetic and potential terms which are coupled to gravity. In the slow-roll limit, when the kinetic term of energy is relatively small, compared to the potential term, the inflation period appears. After this period, in reheating era, the scalar field oscillates around the minimum of the potential losing its energy to light particles \cite{3-m}. Recently, a new model of inflation was proposed \cite{1} and studied \cite{2} which is generally driven by non-Abelian gauge fields. This model is commonplace in particle physics models but is against the isotropic symmetry of the inflationary universe at the background level. This problem could be solved by using three gauge fields which can rotate among each other by non-Abelian gauge transformation SU(2). The rotational symmetry is global part of SU(2) and isotropy is retained in gauge field inflation (gauge-flation) model \cite{1}.\\
In one sector of the present work, we would like to consider gauge-flation model in the context of "intermediate inflation". This scenario is one of the exact solutions of inflationary field equation in the Einstein theory with scale factor $a(t)=a_0\exp(At^f)$ ($A>0, 0<f<1$). The study of this model is motivated by string/M theory \cite{4-m}.  If we add the higher order curvature correction, which is proportional to Gauss-Bonnent (GB) term, to Einstein-Hilbert action then we obtain a free-ghost action \cite{5-m}. The Gauss-Bonnent interaction is leading order of the "$\alpha$" expansion to low-energy string effective action \cite{5-m} ($\alpha$ is inverse string tension). This theory may be applied for black hole solutions \cite{6-m}, acceleration of the late time universe \cite{7-m} and initial singularity problems \cite{8-m}. The GB interaction in $4D$ with dynamical dilatonic scalar coupling leads to an intermediate form of scale factor \cite{4-m}. Expansion of the universe in the intermediate inflation scenario is slower than standard de Sitter inflation with scale factor $a=a_0\exp(H_0 t)$ ($a_0, H_0>0$) which arises as $f=1$, but faster than power-low inflation with scale factor $a=t^p$ ($p>1$). Harrison-Zeldovich \cite{10-m}  spectrum of density perturbation i.e. $n_s=1$ for intermediate inflation model  which is driven by scalar field, is presented for exact values of parameter $f$ \cite{11-m}.\\
On the other hand, we will also study our model in the context of "logamediate inflation" with scale factor $a(t)=a_0 \exp(a[\ln t]^{\lambda})$ ($\lambda >1, A>0$) \cite{12-m}. This model is converted to power-law inflation for $\lambda=1$ cases. This scenario is applied in a number of scalar-tensor theories \cite{13-m}. The study of logamediate scenario is motivated by imposing weak general conditions on the cosmological models which have indefinite expansion \cite{12-m}. The effective potential of the logamediate model has been considered in dark energy models \cite{14-m}. These form of potentials are also used in supergravity, Kaluza-Klein theories and super-string models \cite{13-m,15-m}. For logamediate models the power spectrum could be either red or blue tilted \cite{16-m}. In ref.\cite{12-m}, we could find eight possible asymptotic scale factor solutions for cosmological dynamics. Three of these solutions are non-inflationary scale factor, another three one's of solutions give power-low, de sitter and intermediate scale factors. Finally, two cases of these solutions have asymptotic expansion with logamediate scale factor.

\section{Gauge-flation}
Gauge-flation model in a flat-space Friedmann-Robertson-Walker (FRW) background is described by an effective Lagrangian \cite{1}
\begin{eqnarray}\label{1}
\mathcal{L}=\sqrt{-g}(-\frac{R}{2}-\frac{1}{4}F^{a}_{\mu\nu}F^{\mu\nu}_a+\frac{\kappa^2}{384}(\epsilon^{\mu\nu\lambda\sigma}F^{a}_{\mu\nu}F^{a}_{\lambda\sigma})^2)
\end{eqnarray}
where $8\pi G=M_p^{-2}=1$, $\epsilon^{\mu\nu\rho\sigma}$ is antisymmetric tensor and
\begin{eqnarray}\label{2}
F^{a}_{\mu\nu}=\partial_{\mu}A^{a}_{\nu}-\partial_{\nu}A^{a}_{\mu}-g\epsilon^{a}_{bc}A^{b}_{\nu}A^{c}_{\mu}
\end{eqnarray}
where $\epsilon^{a}_{bc}$ is also antisymmetric tensor.
The contribution of the specific $F^4$ term to the energy-momentum tensor
will take the form of a perfect fluid with equation of state
\begin{eqnarray}\label{}
\nonumber
P=-\rho
\end{eqnarray}
which is used for driving a de Sitter expansion. $F^4$ term has been found in particle physics settings and gauge field theory analysis. In particle physics model, this term may be argued for studying axions in a non-Abelian gauge theory \cite{v}.
It was shown that $F^4$ term can be derived by integrating out an axion field in Chromo-Natural inflation \cite{2}.
Our model can provide a setting for constructing a homogeneous and isotropic inflationary background \cite{1}.
To obtain isotropy symmetry of space-time, the effective inflaton field could be introduced by ansatz \cite{1}
\begin{eqnarray}\label{3}
A^{a}_{\mu}=\{^{\phi(t)\delta^a_i~~~~\mu=i}_{0~~~~~~~~~\mu=0}=\{^{a(t)\psi(t)\delta^a_i~~~\mu=i}_{0~~~~~~~~~~~~\mu=0}=\{^{g^{-2}\dot{a}\sqrt{\gamma(t)}\delta^a_i~~~~~~\mu=i}_{0~~~~~~~~~~~~~~~~~~\mu=0}
\end{eqnarray}
where $a$ is scale factor of spatially flat FRW space-time.
With the above choice, gauge indices are identified with the spatial indices. We actually identify the global part of the gauge group, $SU(2),$ with the rotation group $SO(3)$. When we turn on space components of a vector (\ref{3}), the global part of non-Abelian gauge symmetry (which is the symmetry group of our model (\ref{1})) leads to rotational symmetry in spatial segment of space-time. On the other hands, with the above choice (\ref{3}), the global part of non-Abelian gauge symmetry ($SU(2)$) leads to isotropic space-time or rotational symmetry ($SO(3)$) in space.
From Eqs.(\ref{1}) and (\ref{3}), the reduced effective Lagrangian is presented
\begin{eqnarray}\label{}
\nonumber
\mathcal{L}_{red}=\frac{3}{2}(\frac{\dot{\phi}^2}{a^2}-\frac{g^2\phi^4}{a^4}+\kappa\frac{g^2\phi^2\dot{\phi}^2}{a^6})
\end{eqnarray}
Pressure and energy density of our model have the following forms
\begin{eqnarray}\label{}
\nonumber
P=\frac{\partial(a^3\mathcal{L}_{red})}{\partial a^3}=\frac{1}{3}\rho_{YM}-\rho_{F^4}~~~~\\
\nonumber
\rho=\frac{\partial \mathcal{L}_{red}}{\partial \dot{\phi}}\dot{\phi}-\mathcal{L}_{red}=\rho_{YM}+\rho_{F^4}
\end{eqnarray}
where
\begin{eqnarray}\label{}
\nonumber
\rho_{YM}=\frac{3}{2}(\frac{\dot{\phi}^2}{a^2}+\frac{g^2\phi^4}{a^4})~~~~~\rho_{F^4}=\kappa\frac{g^2\phi^4\dot{\phi}^2}{a^6}
\end{eqnarray}
Recalling the Einstein's equations with the Gauge-flation set up, an important equation is presented
\begin{eqnarray}\label{4}
\dot{H}=-(H^2\psi^2+\dot{\psi}^2+g^2\psi^4)
\end{eqnarray}
Slow-roll parameters of this scenario have been introduced in ref.\cite{1}
\begin{eqnarray}\label{5}
\epsilon=-\frac{\dot{H}}{H^2}\simeq\psi^2(1+\frac{g^4\psi^2}{H^2})=\frac{H^2}{g^4}\gamma(1+\gamma)\\
\nonumber
\eta=-\frac{\ddot{H}}{2H\dot{H}}\simeq\psi^2=\frac{H^2}{g^4}\gamma~~~~~~~~~~~~~~~~~~~~~\\
\nonumber
\delta=-\frac{\dot{\psi}}{H\psi}=\frac{\gamma}{6(\gamma+1)}\epsilon^2~~~~~~~~~~~~~~~~~~~~~~~~
\end{eqnarray}
To have a consistent slow-roll inflation, we demand $\epsilon,\eta,\delta<1$. Where $\delta<1$, the equation (\ref{4}) is reduced to
\begin{eqnarray}\label{6}
\dot{H}=-(H^2\psi^2+g^2\psi^4)=-\frac{H^4}{g^4}\gamma(1+\gamma)
\end{eqnarray}
For gauge-flation model the power spectrum of curvature perturbations is given by
\begin{eqnarray}\label{7}
P_{R}=\frac{1}{8\pi^2\epsilon}(\frac{H}{M_{pl}})^2~\\
\end{eqnarray}
Tensor to scalar ratio, spectral tilt and tensor index are important parameters which could be constrained by observational data \cite{2-m}. These parameters for our model have the following forms \cite{1}
\begin{eqnarray}\label{8}
r=8(P_L+P_R)\epsilon~~~~\\
\nonumber
n_s-1=-2(\epsilon-\eta)\\
\nonumber
n_g=-2\epsilon~~~~~~~~~~~~~
\end{eqnarray}
In the standard models of inflation which are deriven  by scalar field $P_L=P_R$, but for gauge-flation model for very large and very small values of $\gamma$ field we have $P_R\gg P_L$ \cite{1}.
\section{Intermediate inflation}
In this section we will study gauge-flation model in the context of intermediate inflation. Scale factor of intermediate inflation follow the law
\begin{eqnarray}\label{9}
a=a_0\exp(At^f)~~~~~~0<f<1
\end{eqnarray}
where $A$ is positive constant. Using above scale factor the number of e-fold is given by
\begin{eqnarray}\label{10}
N=\int_{t_1}^{t} H dt=A(t^{f}-t_1^{f})
\end{eqnarray}
where $t_1$ is the begining time of inflation. Using Eqs.(\ref{6}) and (\ref{9}), scalar field $\gamma$ and Hubble parameter are found
\begin{eqnarray}\label{11}
\gamma(t)=(-\frac{1}{2}+\frac{1}{2}\sqrt{1+\frac{4g^2(1-f)}{f^3A^3}t^{-3f+2}})\\
\nonumber
H(\gamma)=\alpha(\gamma(\gamma+1))^{\beta}~~~~~~~~~~
\end{eqnarray}
where
\begin{eqnarray}\label{12}
\alpha=fA(\frac{f^3A^3}{g^2(1-f)})^{\beta}~~~~~~~~~~~~~~~\beta=\frac{f-1}{-3f+2}
\end{eqnarray}
Slow-roll parameters $\epsilon$ and $\eta$ in term of field $\gamma$ are derived from Eqs.(\ref{5}),(\ref{9}) and (\ref{11})
\begin{eqnarray}\label{13}
\epsilon=-\frac{\dot{H}}{H^2}=\frac{1-f}{fA}(\frac{f^3A^3}{g^2(1-f)}\gamma(\gamma+1))^{\frac{f}{3f-2}}\\
\nonumber
\eta=-\frac{\ddot{H}}{2\dot{H}H}=\frac{2-f}{2fA}(\frac{f^3A^3}{g^2(1-f)}\gamma(\gamma+1))^{\frac{f}{3f-2}}
\end{eqnarray}
respectively. We present the number of e-folds between two fields $\gamma_1=\gamma(t_1)$ and $\gamma=\gamma(t)$, using Eq.(\ref{10}).
\begin{eqnarray}\label{}
N= A(\frac{(fA)^3}{4g^2(1-f)})^{\frac{f}{-3f+2}}[((2\gamma+1)^2-1)^{\frac{f}{-3f+2}}-((2\gamma_1+1)^2-1)^{\frac{f}{-3f+2}}]
\end{eqnarray}
$\gamma_1$ denotes the begining inflaton ($\epsilon=1$) $\gamma_1=\frac{1}{2}(1+\frac{4g^2}{f^2A^2}(\frac{1-f}{fA})^{\frac{2(1-f)}{f}})-\frac{1}{2}$. Therefor, the value of $\gamma(t)$ could be determined in terms of $N, f, A$
\begin{eqnarray}\label{14}
\gamma=\frac{1}{2}[1+\frac{4g^2(1-f)}{(fA)^3}(\frac{N}{A}+\frac{1-f}{fA})^{\frac{2-3f}{f}}]^{\frac{1}{2}}-\frac{1}{2}
\end{eqnarray}
For intermediate inflation we obtain the perturbation parameters versus the scalar field $\gamma$ and constant parameters of intermediate scenario($f,A$).
From Eq.(\ref{7}), we could find the spectrum of curvature perturbation in slow-roll limit.
\begin{eqnarray}\label{15}
P_R=\frac{1}{8\pi^2M_{pl}^2}\frac{f^3A^3}{1-f}(\frac{(fA)^3}{g^2(1-f)}\frac{(2\gamma+1)^2-1}{4})^{-1}\\
\nonumber
=\frac{1}{8\pi^2M_{pl}^2}\frac{f^3A^3}{1-f}(\frac{N}{A}+\frac{1-f}{fA})^{\frac{3f-2}{f}}
\end{eqnarray}
This parameter may be constrained by WMAP data  \cite{2-m}.
Using Eq.(\ref{8}), tensor-scalar ratio has the form
\begin{eqnarray}\label{16}
r=8(P_L+P_R)\frac{1-f}{fA}(\frac{(fA)^3}{g^2(1-f)}\frac{(2\gamma+1)^2-1}{4})^{\frac{f}{3f-2}}~~~\\
\nonumber
=8(P_L+P_R)\frac{1-f}{fA}(\frac{N}{A}+\frac{1-f}{fA})^{-1}
\end{eqnarray}
In Fig.(1), we plot the tensor-scalar ratio  versus the number of e-folds where $f=\frac{2}{5}$. In this case we find the model is compatible with observational data \cite{2-m},  $N\geq 100$  leads to $0.02<r<0.11$.

\begin{figure}[h]
\centering
  \includegraphics[width=10cm]{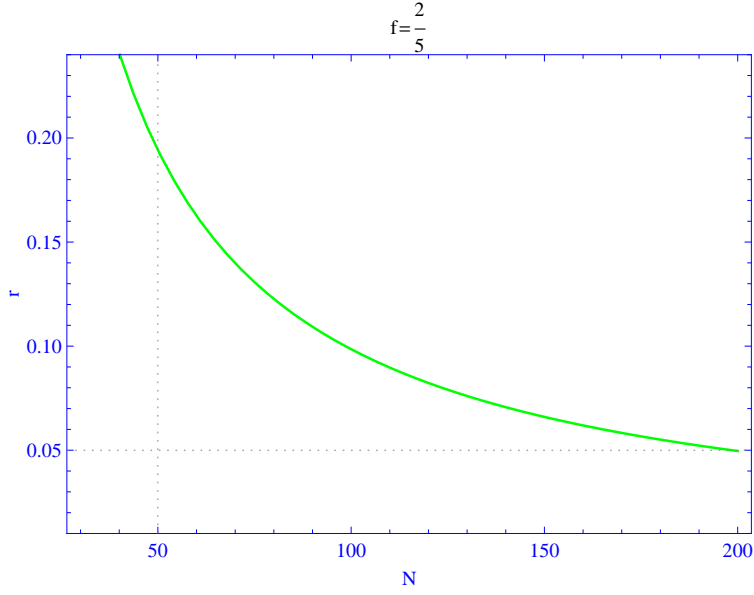}
  \caption{In this graph we plot the tensor-to-scalar ratio  in term of the number of e-folds $N$ for intermediate scenario ($A=1$,$P_L+P_R=6$). }
 \label{fig:F3}
\end{figure}
This parameter could be constrained by  Planck satellite data ($0.02<r<0.11$) \cite{2-m}.
Finally spectral and tensor indices have following forms
\begin{eqnarray}\label{17}
n_s-1=\frac{d\ln P_R}{dN}=\frac{3f-2}{fA}(\frac{N}{A}+\frac{1-f}{fA})^{-1}~~~~~~~~~~~~~~~~~~~~~~\\
\nonumber
n_T=-2\frac{1-f}{fA}(\frac{(fA)^3}{g^2(1-f)}\gamma(\gamma+1))^{\frac{f}{3f-2}}=\frac{2f-2}{fA}(\frac{N}{A}+\frac{1-f}{fA})^{-1}
\end{eqnarray}
Spectral index $n_s$ in term of number of e-folds is plotted in Fig.(2). Standard case $N\simeq 50$, leads to $n_s\simeq0.96$ which is agree with observational data \cite{2-m}.
\begin{figure}[h]
\centering
  \includegraphics[width=10cm]{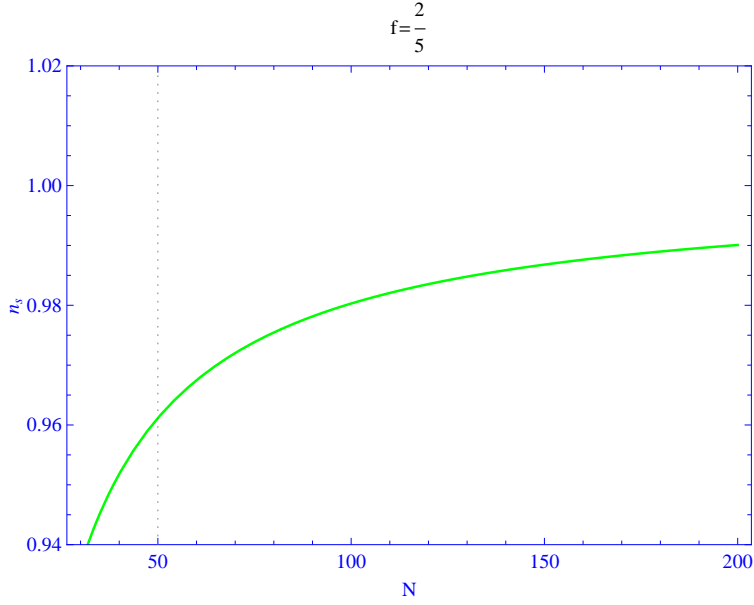}
  \caption{In this graph we plot the spectral index $n_s$   in term of the number of e-folds $N$ for intermediate scenario ($A=1$).}
 \label{fig:F3}
\end{figure}
\section{Logamediate inflation}
Now, we consider Gauge field logamediate inflation. In logamediate scenario the scale factor has the following form
\begin{eqnarray}\label{18}
a(t)=a_0\exp(A[\ln t]^{\lambda}) ~~~~~~\lambda>1,
\end{eqnarray}
$A$ is a constant parameter. From this scale factor, we find the number of e-folds
\begin{eqnarray}\label{19}
N=\int_{t1}^{t} H dt=A[(\ln t)^{\lambda}-(\ln t_1)^{\lambda}]
\end{eqnarray}
where $t_1$ denotes the begining of inflation period. Using Eqs.(\ref{6}) and (\ref{18}), we get the field $\gamma$ at late time.
\begin{eqnarray}\label{20}
\gamma=\Xi(t)
\end{eqnarray}
where $\Xi(t)=-\frac{1}{2}+\frac{1}{2}\sqrt{1+\frac{4g^2t^2}{(\lambda A)^3(\ln t)^{3(\lambda-1)}}}$ and
\begin{eqnarray}\label{21}
\dot{H}=\frac{\lambda A(\ln t)^{\lambda-1}}{t^2}[\frac{\lambda-1}{\ln t}-1]
\end{eqnarray}
First term in the above equation may be omitted at the late time. Slow-roll parameters $\epsilon$ and $\eta$ at the late time are given by
\begin{eqnarray}\label{22}
\epsilon=\frac{(\ln \Xi^{-1}(\gamma))^{1-\lambda}}{\lambda A}~~~~~~~~~~~~~~~~~~~~~\\
\eta=\frac{(\ln \Xi^{-1}(\gamma))^{1-\lambda}}{\lambda A}[1-\frac{\lambda-1}{2\ln \Xi^{-1}(\gamma)}]
\end{eqnarray}
$\Xi^{-1}$ is inverse function of $\Xi$.
We derive the Hubble parameter in term of scalar field $\gamma$
\begin{eqnarray}\label{23}
H=\frac{A\lambda[\ln \Xi^{-1}(\gamma)]^{\lambda-1}}{\Xi^{-1}(\gamma)}
\end{eqnarray}
The number of e-folds between two fields $\gamma_1=\gamma(t_1)$, $\gamma=\gamma(t)$ could be determined from Eq.(\ref{19}).
\begin{eqnarray}\label{24}
N=\lambda [(\ln \Xi^{-1}(\gamma))^{\lambda}-(\ln \Xi^{-1}(\gamma_1))^{\lambda}]~~~~~~~~~\gamma=\Xi(\exp(\frac{N}{A}+(\lambda A)^{\frac{\lambda}{1-\lambda}}))
\end{eqnarray}
$\gamma_1$ is found at the begining of the inflation where $\epsilon=1$
\begin{eqnarray}\label{25}
\gamma_1=\Xi(\exp([\lambda A]^{\frac{1}{1-\lambda}}))
\end{eqnarray}
Perturbation parameters of the model for logamediate scenario are obtained from Eq.(\ref{15}),(\ref{16})and (\ref{17}).
In slow-roll limit, spectrum of curvature perturbation is given by
\begin{eqnarray}\label{26}
P_R=\frac{(\lambda A)^3}{8\pi^2M_{pl}^2}\frac{(\ln\Xi^{-1}(\gamma))^{3\lambda-3}}{(\Xi^{-1}(\gamma))^2}~~~~~~~~~~~~~~~~~~~~~~~~~~~~~~~\\
\nonumber
=\frac{(\lambda A)^3}{8\pi^2M_{pl}^2}\exp(-2(\frac{N}{A}+(\lambda A)^{\frac{\lambda}{1-\lambda}}))(\frac{N}{A}+(\lambda A)^{\frac{\lambda}{1-\lambda}})^{3\lambda-3}
\end{eqnarray}
Spectral and tensor indices of this model are derived from Eq.(\ref{8})
\begin{eqnarray}\label{27}
n_s=1-\frac{\lambda-1}{\lambda A}(\ln \Xi^{-1})^{-\lambda}=1-\frac{\lambda-1}{\lambda A}(\frac{N}{A}+(\lambda A)^{\frac{\lambda}{1-\lambda}})^{-\lambda}\\
\nonumber
n_T=-2\frac{(\ln \Xi^{-1}(\gamma))^{(1-\lambda)}}{\lambda A}~~~~~~~~~~~~~~~~~~~~~~~~~~~~~~
\end{eqnarray}
Spectral index $n_s$ in term of number of e-folds is plotted in Fig.(3) (for $\lambda=2$, $\lambda=3$, $\lambda=4$ and $\lambda=5$ cases). It is observed that small values of number of e-folds are assured for large values of $\lambda$ parameter.
\begin{figure}[h]
\begin{minipage}[b]{1\textwidth}
\subfigure[\label{fig1a} ]{ \includegraphics[width=.37\textwidth]%
{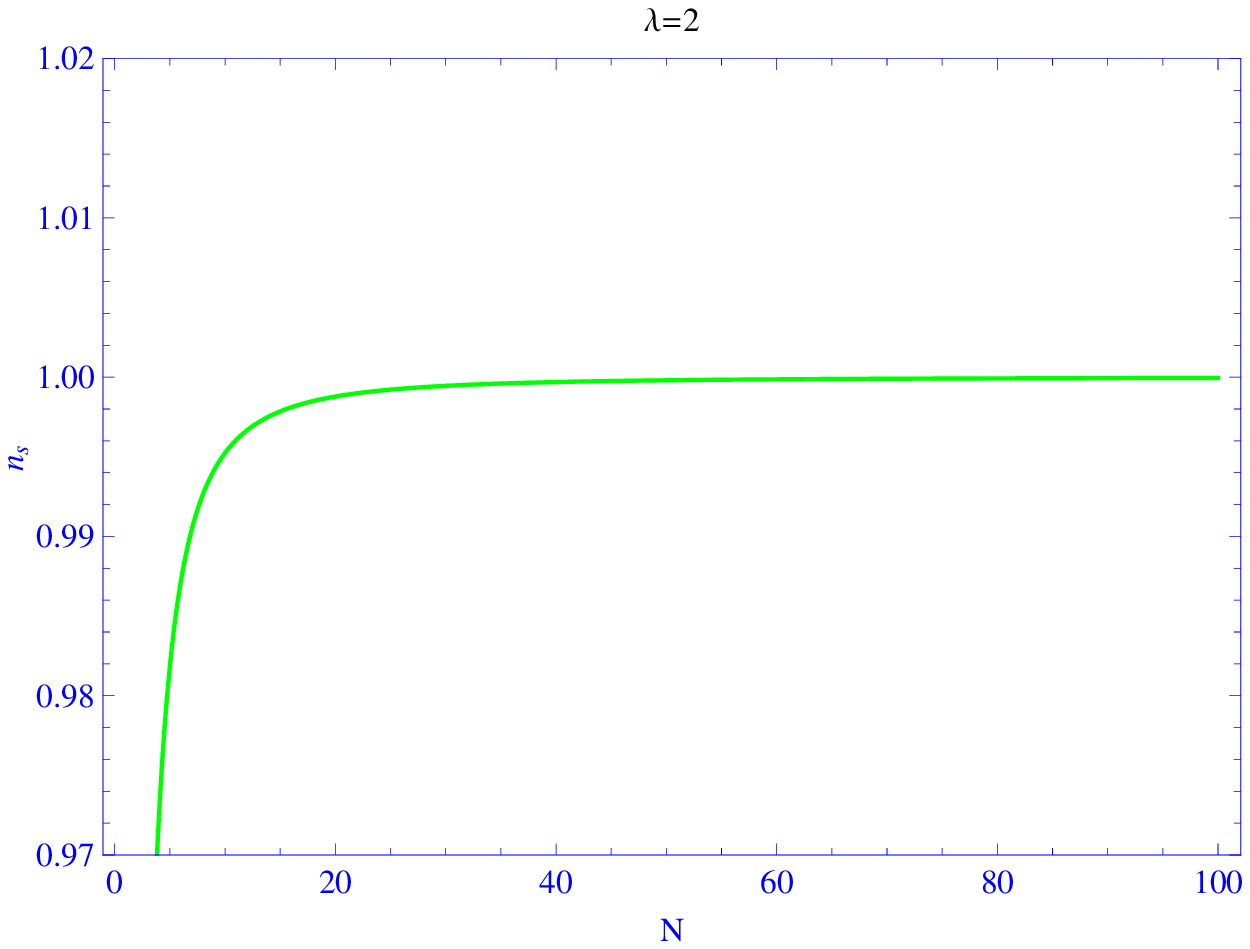}} \hspace{.2cm}
\subfigure[\label{fig1b} ]{ \includegraphics[width=.37\textwidth]%
{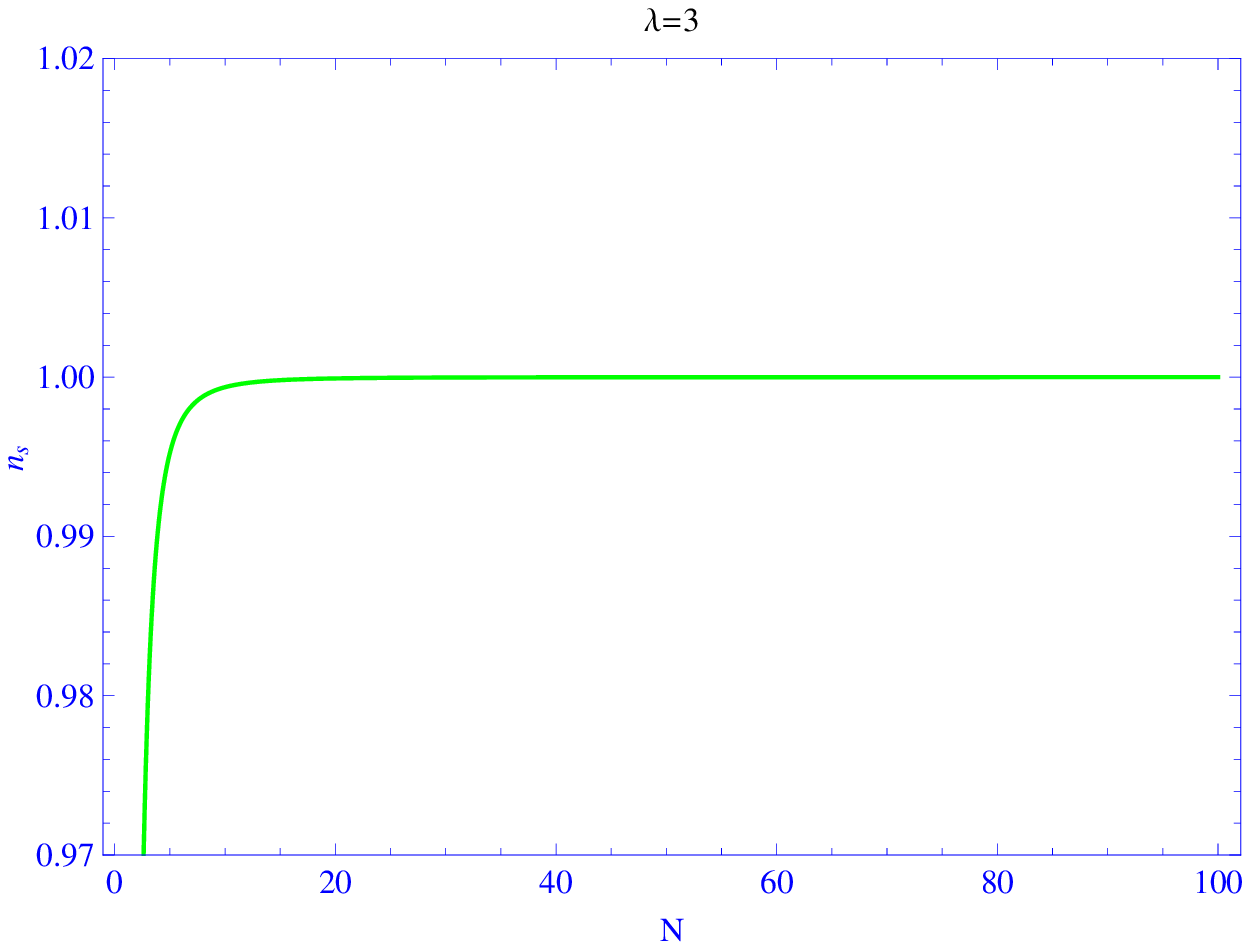}}
\subfigure[\label{fig1a} ]{ \includegraphics[width=.37\textwidth]%
{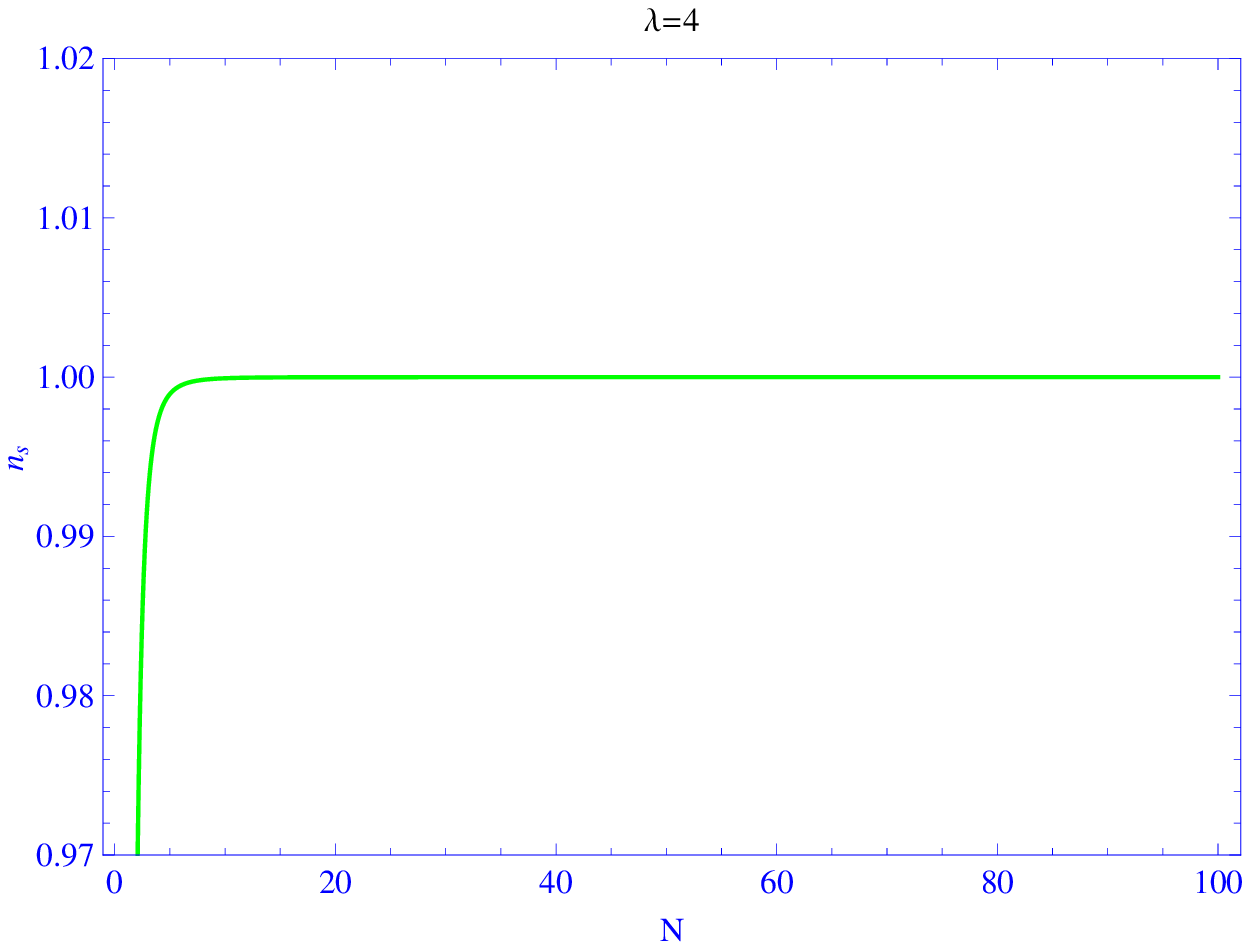}} \hspace{.2cm}
\subfigure[\label{fig1a} ]{ \includegraphics[width=.37\textwidth]%
{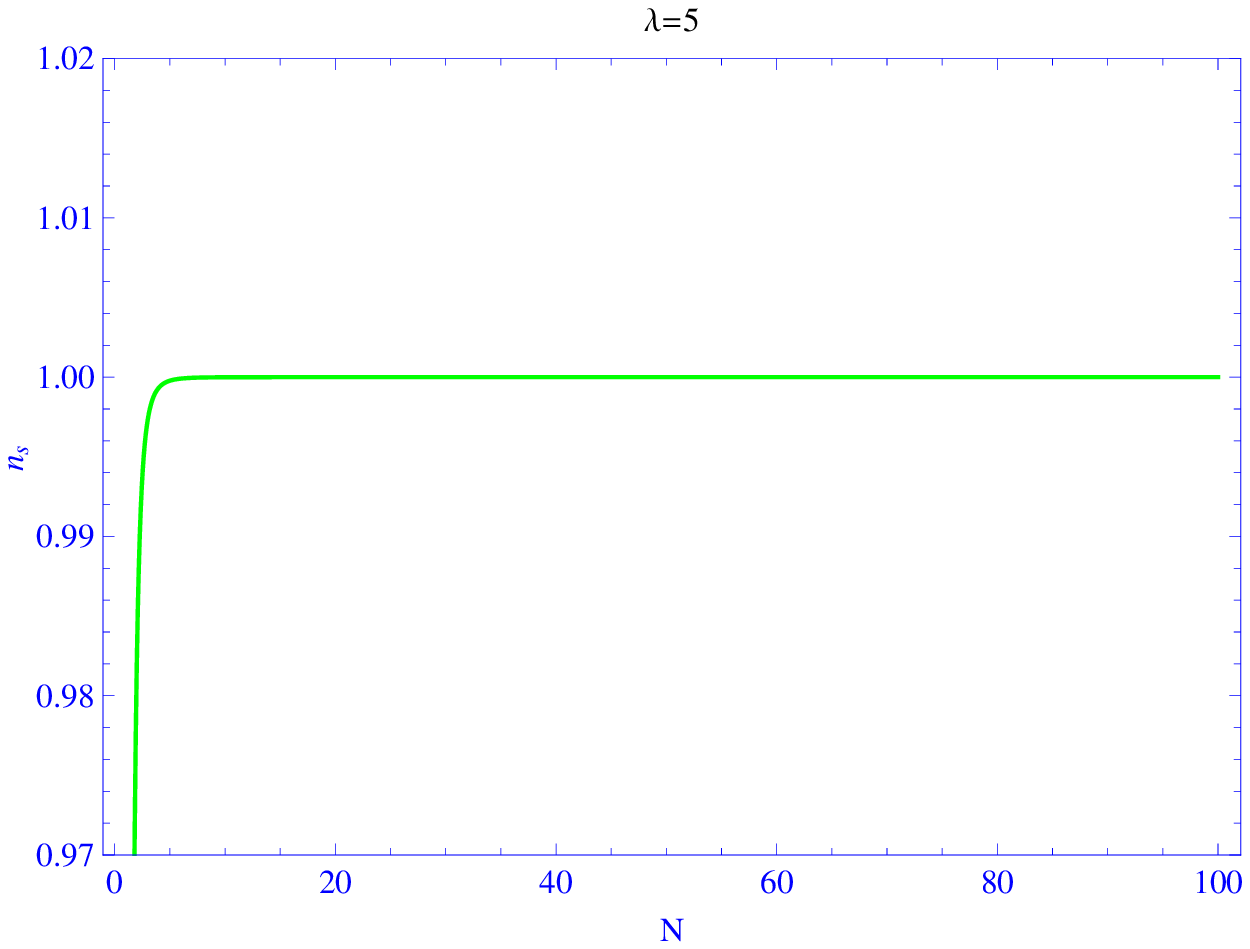}} \hspace{.2cm}
\end{minipage}
\caption{ Spectral index $n_s$ in term of number of e-folds $N$: (a) for $\lambda=2$ and (b) for $\lambda=3$ (c) for $\lambda=4$ (d) for $\lambda=5 $(where   $A=1$ and $P_L+P_R=6$).}
\end{figure}
Tensor-to-scalar ratio for Gauge-flation model in logamediate scenario has the form
\begin{eqnarray}\label{28}
r=8(P_L+P_R)\frac{(\ln \Xi^{-1}(\gamma))^{1-\lambda}}{\lambda A}~~~~~~~~~\\
\nonumber
=\frac{8(P_L+P_R)}{\lambda A}(\frac{N}{A}+(\lambda A)^{\frac{\lambda}{1-\lambda}})^{1-\lambda}
\end{eqnarray}
In Fig.(4), we plot the tensor-scalar ratio in versus the number of e-folds where for $\lambda=2$, $\lambda=3$, $\lambda=4$ and $\lambda=5$. In $\lambda=2,$ case we find the model is compatible with observational data \cite{2-m}.
\begin{figure}[h]
\begin{minipage}[b]{1\textwidth}
\subfigure[\label{fig1a} ]{ \includegraphics[width=.37\textwidth]%
{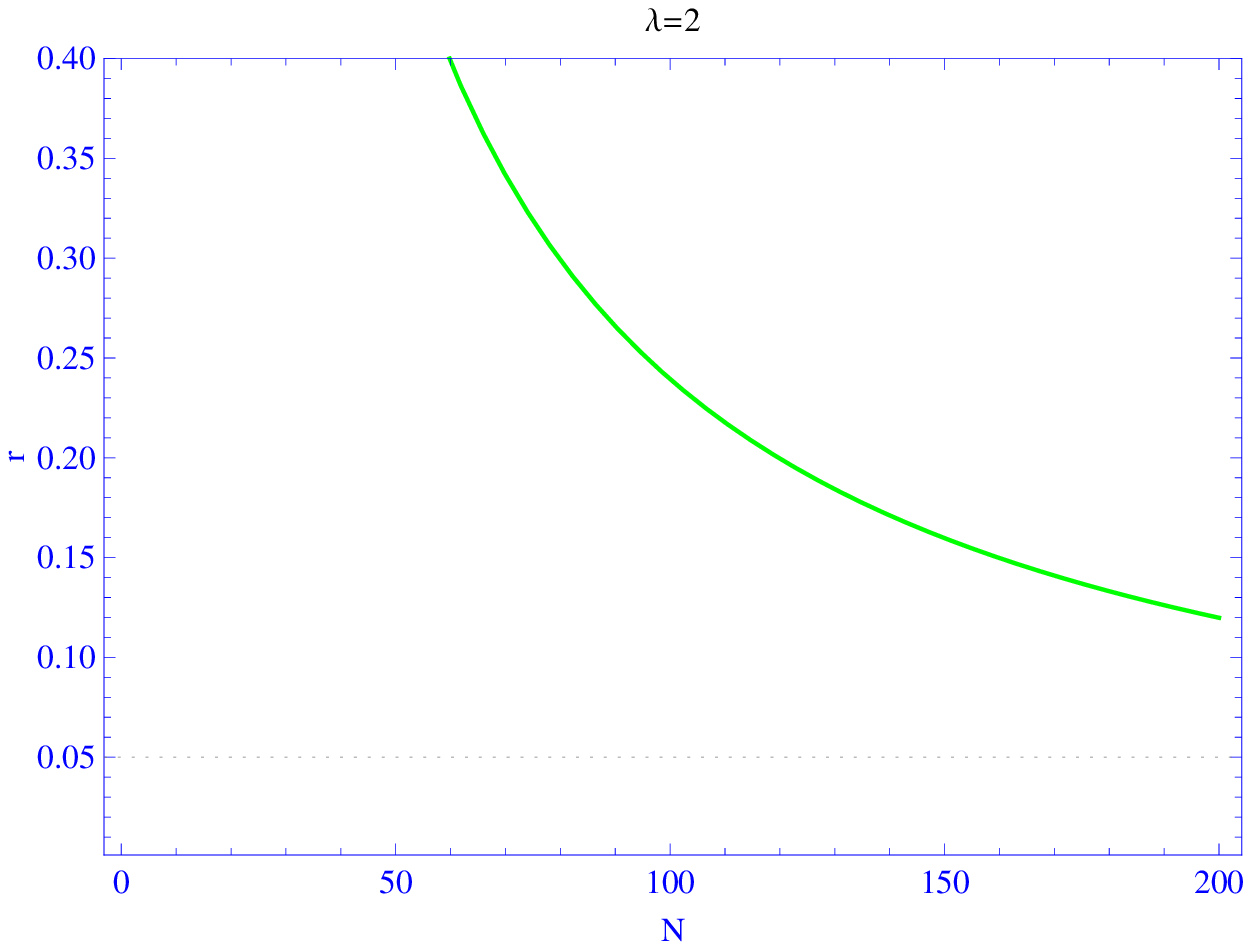}} \hspace{.2cm}
\subfigure[\label{fig1b} ]{ \includegraphics[width=.37\textwidth]%
{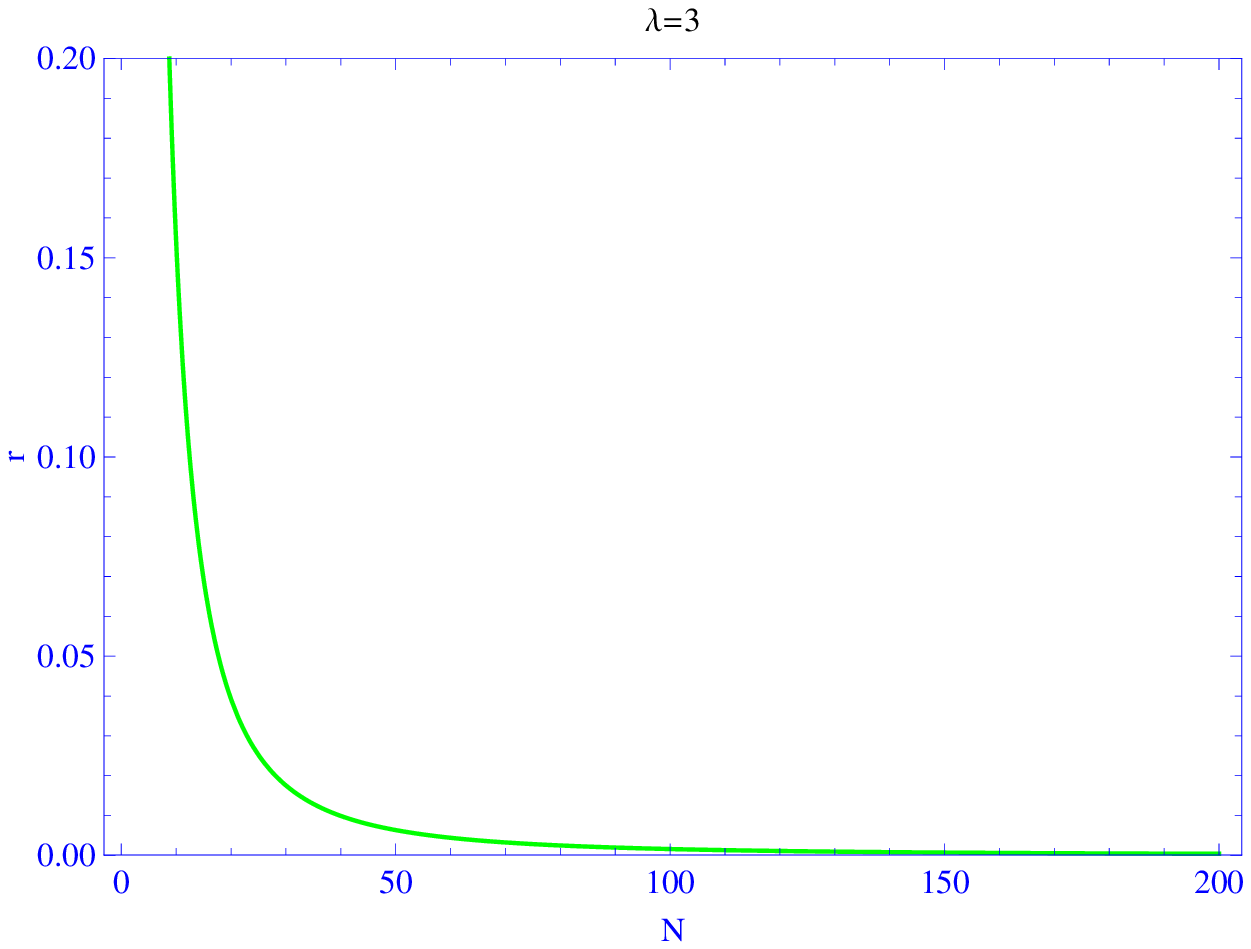}}
\subfigure[\label{fig1a} ]{ \includegraphics[width=.37\textwidth]%
{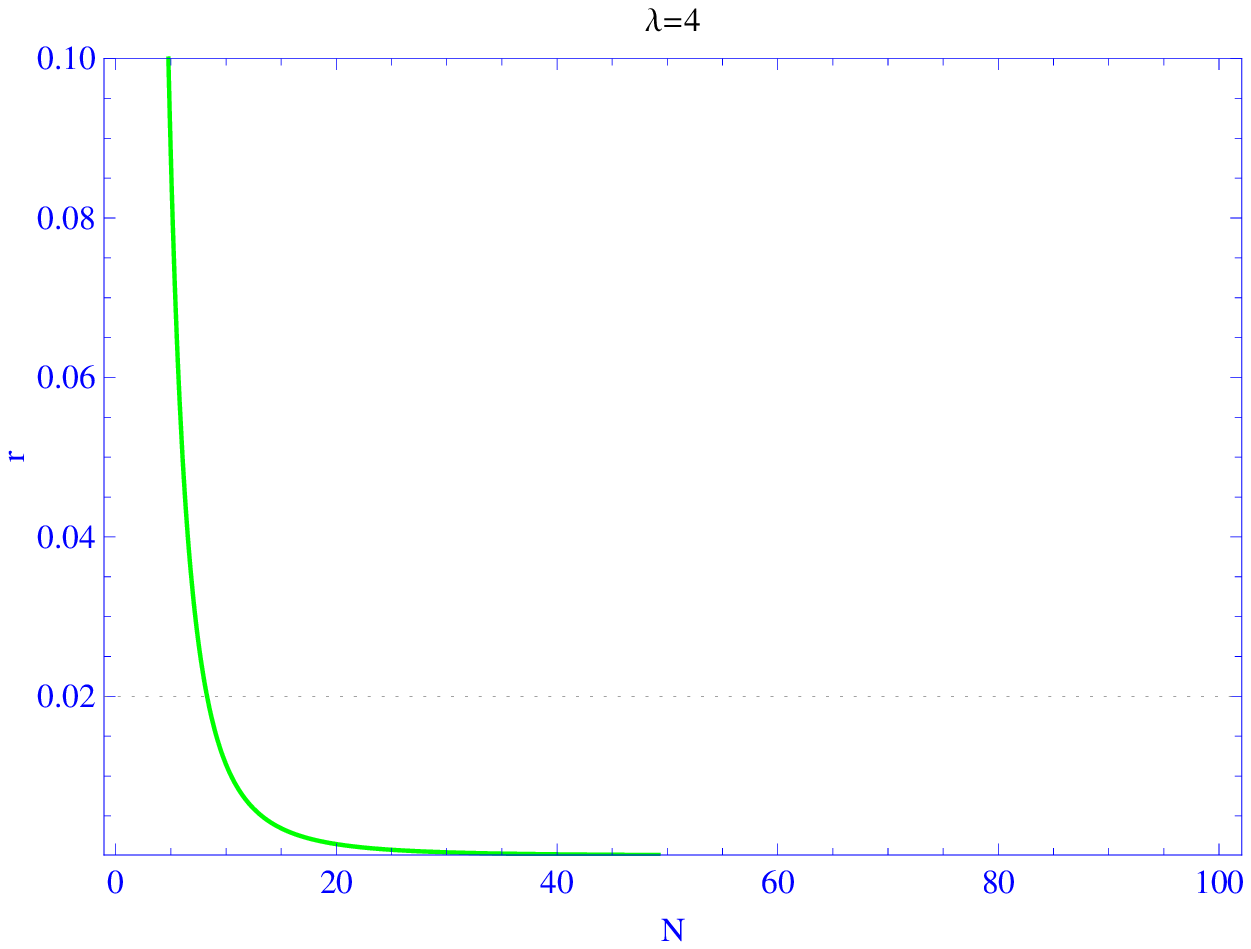}} \hspace{.2cm}
\subfigure[\label{fig1a} ]{ \includegraphics[width=.37\textwidth]%
{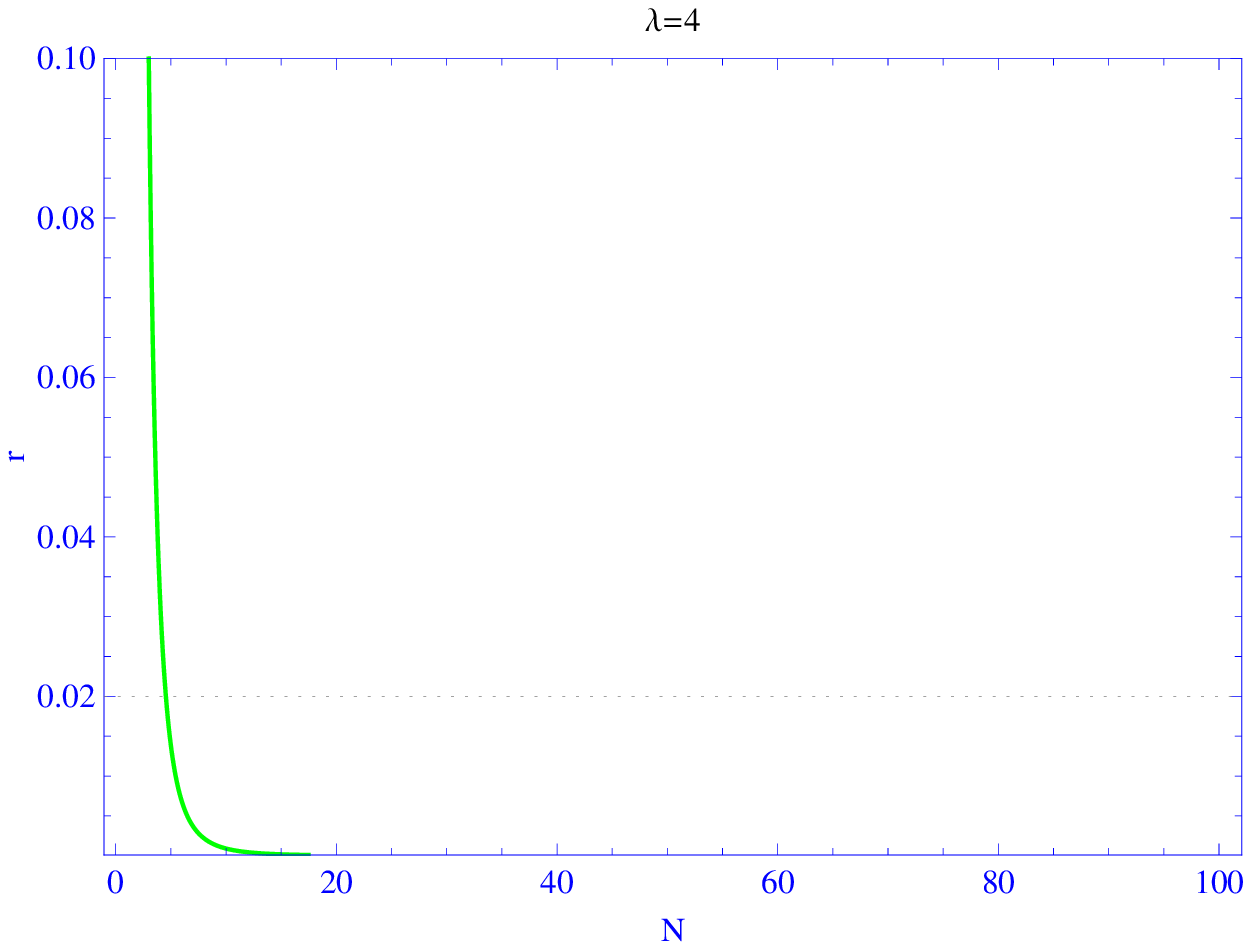}} \hspace{.2cm}
\end{minipage}
\caption{ Tensor-scalar ratio  in term of number of e-folds  $N$: (a) for $\lambda=2$ and (b) for $\lambda=3$ (c) for $\lambda=4$ (d) for $\lambda=5 $(where   $A=1$ and $P_L+P_R=6$).}
\end{figure}
\section{Conclusion}
In this paper we have studied inflation model driven by non-abelian gauge fields.
The action of this model are found in particle physics high energy models. In one segment of the article,
the model has been considered in the context of intermediate inflation. In this scenario we have found perturbation parameters $P_R$, $r$ and $n_s$ in term of inflaton field. We also have connected these parameters to observational data. Our model is compatible with WMAP9 and Planck data. On the other hand we have studied gauge-flation model using Logamediate scenario. Hubble parameter, slow-roll parameters and perturbation parameters of this model are presented in Logamediate inflation. In this case we have found that $\lambda=2$ is compatible with WMAP9 data.


\end{document}